# LOADng Routing Protocol Evaluation for Bidirectional Data flow in AMI Mesh Networks

Saida ELYENGUI[1], Riadh BOUHOUCHI[2], Tahar EZZEDINE[3]

[1.2.3]*Communication System Laboratory Sys'Com, National Engineering School of Tunis, University Tunis El Manar*

***Abstract*— This research work denotes a novel evaluation of LOADng the routing protocol for Wireless Sensor Network. The LOADng protocol implementation is a part of ITU-T G.9903 recommendation based on the framework of the LLN On-demand Ad hoc Distance-vector Routing Protocol - Next Generation (LOADng) proposed by IETF specified by the IETF Internet-Draft draft-clausen-lln-loadng-11 and currently still in its design phase. LOADng is a reactive on demand distance vector routing protocol derived from AODV the Ad hoc On-demand distance vector protocol proposed by IETF. This work was motivated by the need for a novel protocol implementation for smart metering applications providing better performance and less complexity than RPL the Routing Protocol for Low power and lossy networks and adapted to (LLNs) requirements and constraints.**

**Our implementation was successfully integrated into the communication layer of Contiki OS the Wireless Sensor Network operating system and evaluated through extensive simulations for AMI Mesh Networks.**

## I. INTRODUCTION

Smart metering is expected to be an integral part of the smart grid, since advanced metering infrastructure is the foundation of the power grid, providing the connection between customer's premises and neighbor area networks (NAN) in order to transport metering and configuration data with bidirectional traffic flow to and from the information systems of energy providers.

The main purpose of automated metering is to enable real time access to metering data in order to improve management and production of energy and the configuration of smart meters in real time [12].

In this context, one of the prime challenges is providing scalable communication for bidirectional data flow to collect and manage large amount of data for distribution domain. Therefore the choice of a suitable routing protocol providing robust and scalable performance for different types of data traffic is mandatory.

This paper describe the evaluation of a novel routing protocol LOADng [1] which is still in the design phase by IETF to realize smart metering communications.

Results from extensive simulations carried out on cooja the contiki OS simulator on a realistic network topology are presented to demonstrate the efficiency of our proposed solution on the network latency, Packet delivery Ratio and control traffic overhead.

In this manuscript we make the following contributions:

We evaluate our implementation of LOADng protocol and compare it to AODV protocol and RPL protocol for bidirectional scenarios in AMI mesh networks architecture.

We provide analytical results for the network end-to-end delay, PDR, overhead and show how our implemented LOADng solution can improve bidirectional data flow scalability.

The paper is organized as follows: Section II provides an overview about LOADng protocol for AMI mesh networks and highlights the prime issues raised that motivated us to its implementation.

Section III provides a detailed overview of LOADng protocol. Section IV presents detailed performance evaluation and analytical results, in section V we conclude.

## II. LOADNG PROTOCOL

The LLN on-demand ad hoc distance vector routing protocol – next generation LOADng is a reactive routing protocol for Wireless Sensor Networks. It is a simplified version of ad hoc on-demand routing protocol AODV originally developed for use in IEEE 802.15.4 based devices in 6LoWPANs and LLNs [2]. This protocol may be used at layer 3 as a route-over routing protocol or at layer 2 as a mesh-under protocol. Therefore, LOADng algorithm is characterized by its simplicity and its low memory storage needs. Thus, it would be ideal and suitable solution for AMI mesh networks [9, 10, 11]. As it was originally developed to WSNs and Low Power and Lossy Networks (LLNs), it should be adapted to their requirements and constraints.

### *A. Motivation*

Mesh network for smart metering application can be considered as trees of nodes rooted to different concentrators creating a neighbor area networks.

Each concentrator is serving tree of smart meters in the same neighborhood collecting data and transmitting it threw different gateways to the utility information systems. Such scenario requires a bidirectional dataflow between nodes and concentrator in frequent way for real time metering and energy load management.

Different routing protocols were developed for LLNs [4] the most famous is RPL the routing protocol for low power and lossy networks standardized by IETF [5, 13, 14] supporting and optimized for most of LLNs traditional scenarios where MP2P and P2P communication traffic patterns are the most frequent and where traffic from concentrator to sensors are rare occurrence. Whereas, other P2MP scenarios should be also considered for study in AMI mesh networks. Where appropriate and predominant bidirectional dataflow scenarios are MP2P and P2MP traffic patterns. So our motivation for LOADng implementation came from the need for a novel protocol suitable for AMI.

### *B. LOADng Specification*

LOADng describes four types of packets:

- Route Request (RREQ): The RREQ is generated by a router the <originator>, when a data packet in available to a destination, RREQ packet is with no valid route and with a specific destination address.
- Route Reply (RREP): The RREP is generated by a router, upon a RREQ reception and processing with destination address in its routing set.
- Route Reply Acknowledgement (RREP-ACK): The RREP-ACK is generated by a LOADng router after a reception of RREP, as an indication to the neighbor source of the RREP that the RREP was successfully received.
- Route Error (RERR): the RERR is generated by a router when the router detects a broken route to the destination.

LOADng inherited basic operations of AODV, including generation and forwarding of Route Request RREQs to discover a route to a specific destination as shown in Fig 1.

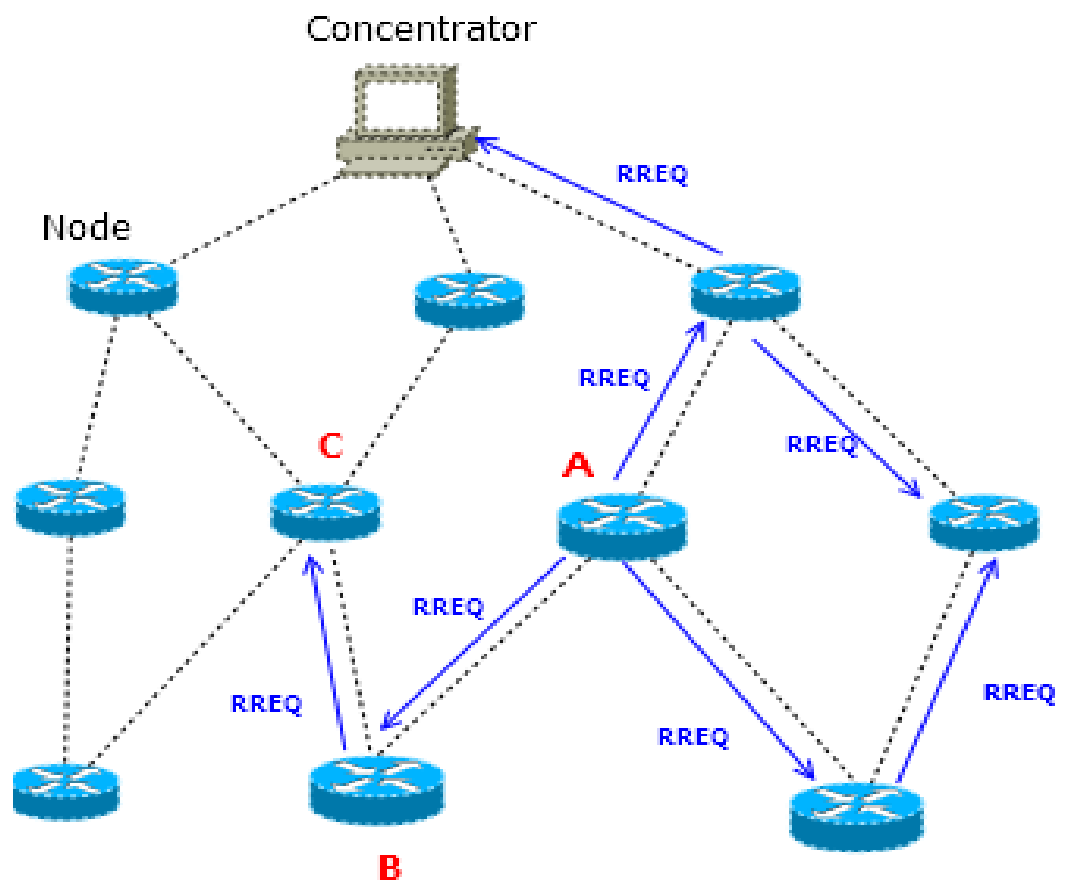

**Fig. 1. RREQ forwarding in LOADng**

Upon receiving this message, only the terminator (Node C) can reply by a RREP and forward it on unicast, hop-by-hop to the source as detailed in Fig 2.

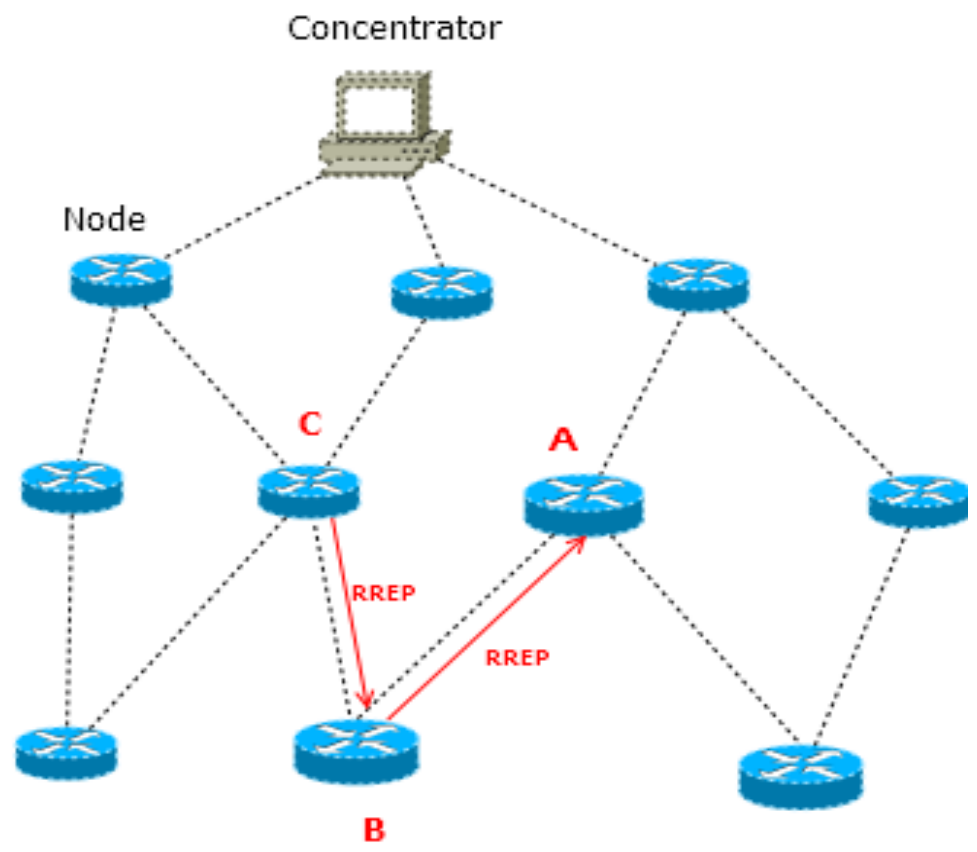

**Fig. 2. RREP unicast forwarding in LOADng.**

When intermediate nodes receive the RREP, they will unicast a proper RREP–ACK to the neighbor from which they received the RREP, in order to notify that the link is bidirectional. If a route is detected broken, an error message can be returned to the source of that data packet.

Compared to AODV, intermediate nodes are not allowed to generate RREPs. As a result, LOADng reduces the size of control messages which is demanded in LLNs

LOADng uses an alternative discovery scheme, denoted Smart RREQ; when receiving a RREQ, intermediate node looks in its routing table entries, if the demanded address exists and it is the next-hop, the RREQ is unicast to the next-hop instead of being broadcast.

In LOADng routing discovery process, only the destination is permitted to respond with a RREP when it receives RREQ message with the same IP address. So, there is no more need to the sequence Number included in AODV messages sent to requesting routers.

Also, there is no more Gratuitous RREP; when an intermediate node has a usable route to the destination the router responds with a RREP on unicast to the source and notify the destination with this Gratuitous message.

Thus, message size would be reduced which is definitely suitable to LLNs low-power and memory constraints [3].

In the other hand, nodes with LOADng protocol do not maintain a precursor list having the IP address for neighbors containing a next hop towards each destination – as it is done with AODV protocol –, but they only care about the next hop to forward current packet to its destination.

LOADng Control messages can include TLV (Type-Length-Value) elements, permitting protocol extensions to be developed [3].

### C. Integration into the Contiki OS

As we work with Contiki OS [8] and the modular Rime stack, we have replaced its main modules which are route and route discovery.

When there is data to be sent, LOADng router discover a bi-directional route to any required destination in the network using the route discovery module. And maintain an active route as long as there is traffic to be sent with the protocol operations implemented in the route module. Each node that we call “router” has a number of parameters, an information base and can generate process and forward a message.

### D. Information Base

In order to maintain the protocol state, the following information base sets are required:

- The “Routing Set”: The Routing Set stores tuples for each reachable node.
- The “Destination Address Set”: Destination Address Set records address for which the current router must respond with a RREP message.
- The “Blacklist Neighbor Set: Blacklist neighbor set stores neighbors to which connectivity is unidirectional.
- The “Pending Acknowledgment Set”: that Contains information about transmitted RREP with a RREP-requirement Set.
- Local Interface Set: Records local LOADng interfaces.

### E. Route module : *“route.c” & route.h*

Route module is the core of the router; it governs the routing table by updating, removing, adding and maintaining routes. In this module, we modified two files; the route protocol route.c and the route protocol header route.h. This module was correctly compiled and integrated into the Contiki OS Rime Stack.

### F. Route-discovery module: *route-discovery.c & route-dicovery.h*

Route-discovery module with route module permit to discover new routes and update routing sets, it uses mesh and uip-over-mesh libraries to enable routers to send and receive messages.

In our case, it deals especially with rime module -using rime address- in order to achieve these functions.

In this module, two files where modified; the route protocol code route-discovery.c and the route protocol header route-dicovery.h. The standard route discovery module of contiki OS was replaced by our modified module with respect to the specification in [1].

## III. Performance Evaluation

### A. Simulations configuration

In order to understand both the performance of LOADng and the performance impact of our implementation we evaluate the LOADng routing protocol in terms of packet Delivery ratio (PDR), latency in order to predict how it behave in larger networks, and Overhead to describe its power consumption and memory management. Simulations were completed in a field of 1000 × 1000 meters, with variable amounts of routers positioned randomly as detailed in table 2.

**Table 1**
**CONTIKI OS AND COOJA SIMULATOR PARAMETER SETUP**

| Settings Transport layer UDP | Value |
|---|---|
| Wireless channel model | UDGM Model with Distance Loss |
| Communication range | 150m |
| Distance to the Concentrator | Variable [20-250] Meters |
| Grid Size | 1000*1000 m2 |
| Number of routers | Variable [25/50/75/100/125/150] |
| Mote type | Tmote Sky |
| Network layer | µIPv6 + 6LoWPAN |
| MAC layer | CSMA + ContikiMAC |
| Radio interface | CC2420 2.4 GHz IEEE 802.15.4 |
| Simulation time | 8h |

The network scenarios are substance of two different traffic patterns: multipoint-to-point (MP2P), wherever all routers generate CBR traffic stream by periodic reporting of 512-byte data packet with 60 seconds interval and acknowledgment of each received frame in upward direction, for which the destination is always the sink.

**TABLE 2**
**TRAFFIC PATTERN OF NODES**

| Node Type | Traffic Pattern |
|---|---|
| Client | MP2P traffic flow by periodic reporting with 5s interval and acknowledgment of each received frame in upward direction. |
| Sink | P2MP traffic with two messages types : Acknowledgment of data frames in downward direction every data arrival. Configuration data sent with Poisson process with average of a single arrival per interval of 10 minutes in downward direction. |

And point-to- multipoint (P2MP) traffic with two messages types, acknowledgment of data frames in downward direction every data arrival and configuration data sent with Poisson process through average of a single arrival per interval of 10 minutes in downward direction. We used two types; client's nodes and a single router node.

*B. Simulations Results*

- *Point-to-multipoint (P2MP):* Fig.4. shows the average packet delivery ratios function of variable distance to the concentrator and Fig.5. shows PDR function of variable nodes number, incurring from respectively AODV, LOADng and RPL.

Fig.6. depicts the delivery ratio for P2MP traffic flow of three protocols; the ratios are close to 100%, regardless of number of nodes. Whereas according to Fig.4. LOADng performance decrease when the node distance to the concentrator is higher than 150 meters. However LOADng yields 60% higher data delivery ratios than does AODV.

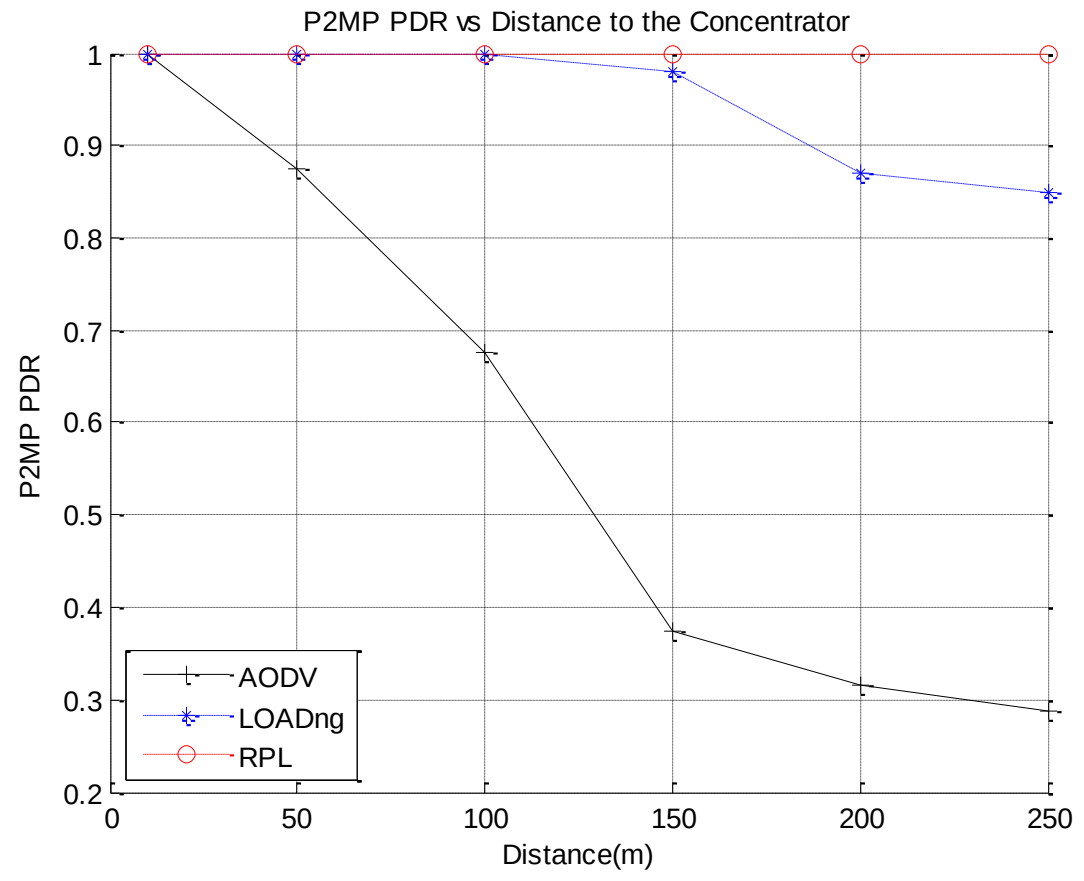

**Fig. 3. Fig.4. P2MP PDR function of Distance**

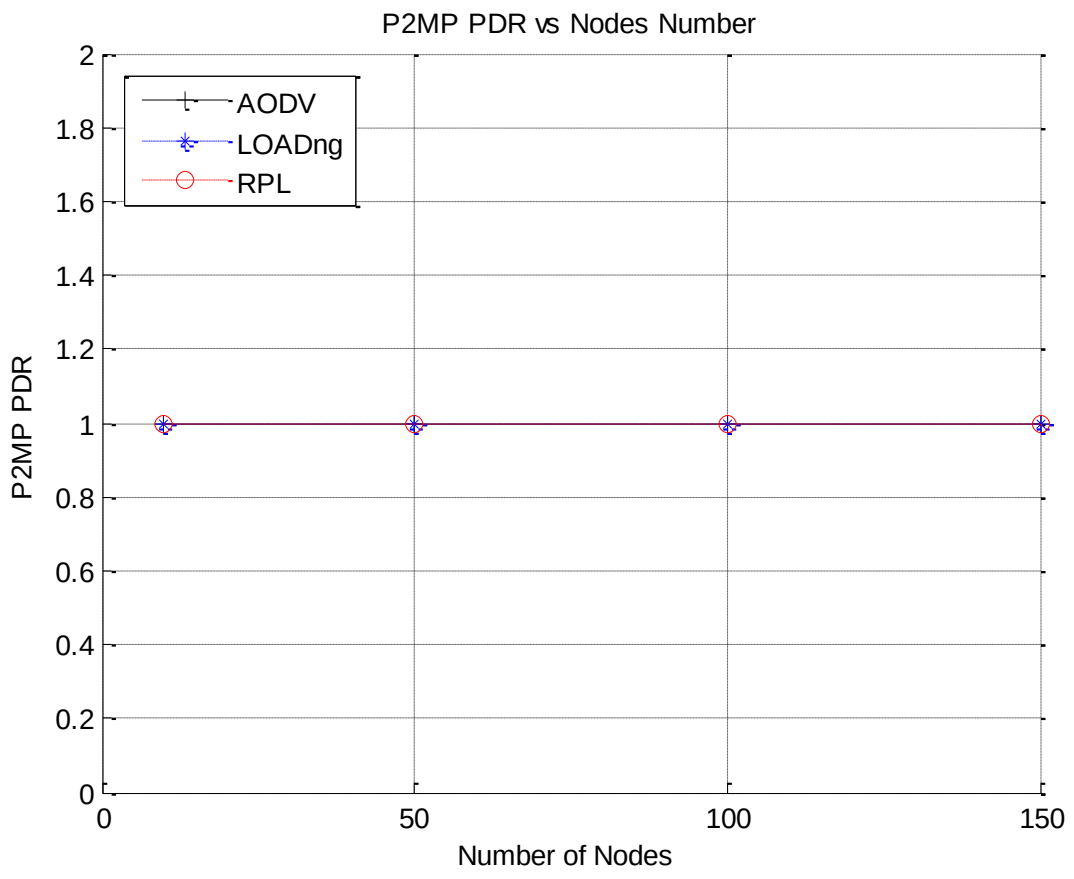

*P2MP PDR function of Nodes Number*

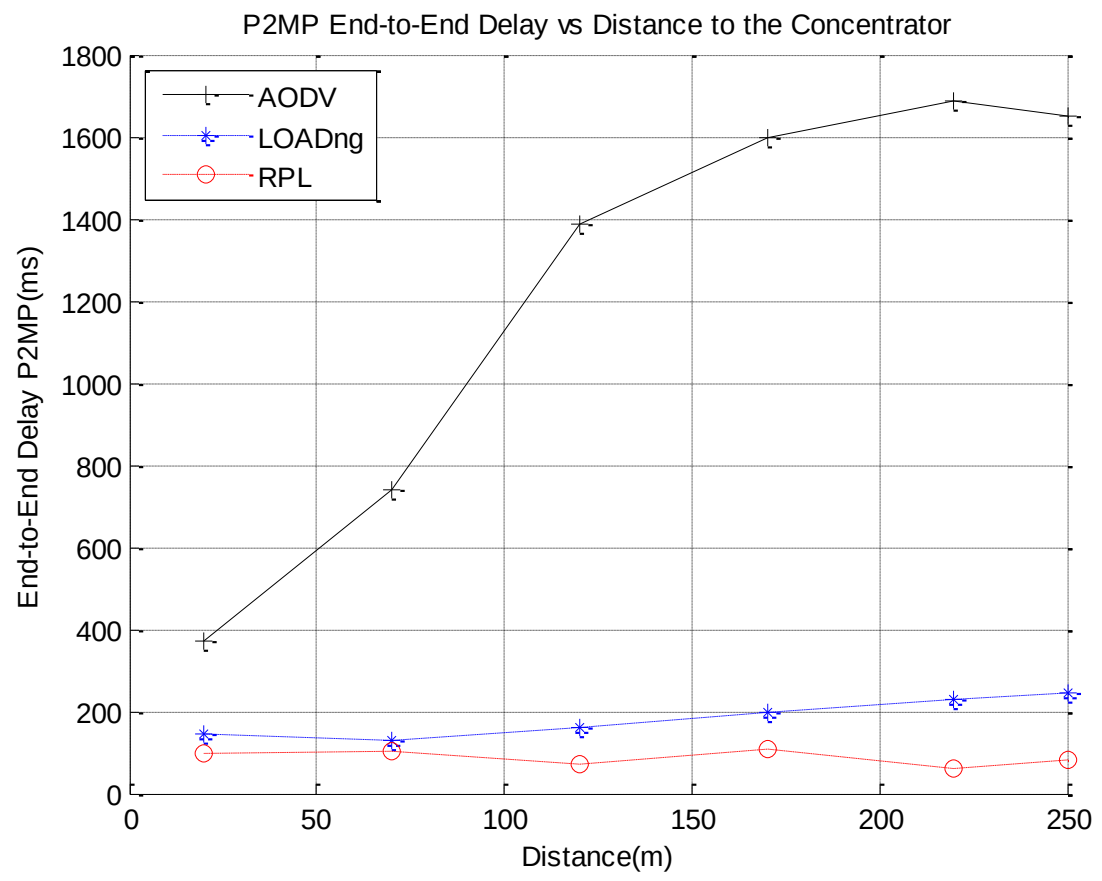

**Fig. 4. P2MP End-to-End Delay function of Distance**

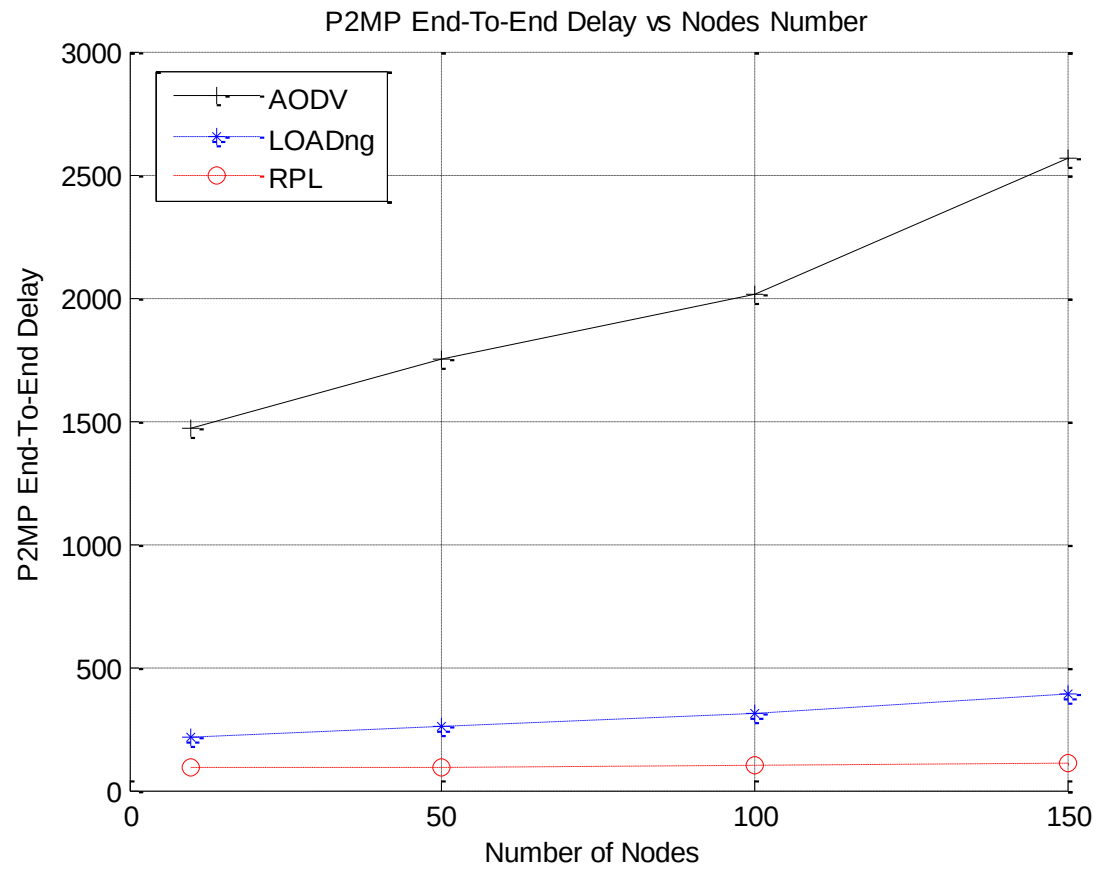

**Fig. 5. P2MP End-to-End Delay function of Node Number**

- *Multipoint-to- Point (MP2P):*

For downward traffic, the Fig.8 shows that LOADng is very efficient in terms of average latency which is equal to 286 ms compared to AODV 1680 ms when the network is subject to variable distance to the concentrator and when the network is with increasing number of nodes is variable latency is equal to 463 ms and 2586 ms for LOADng and AODV respectively.

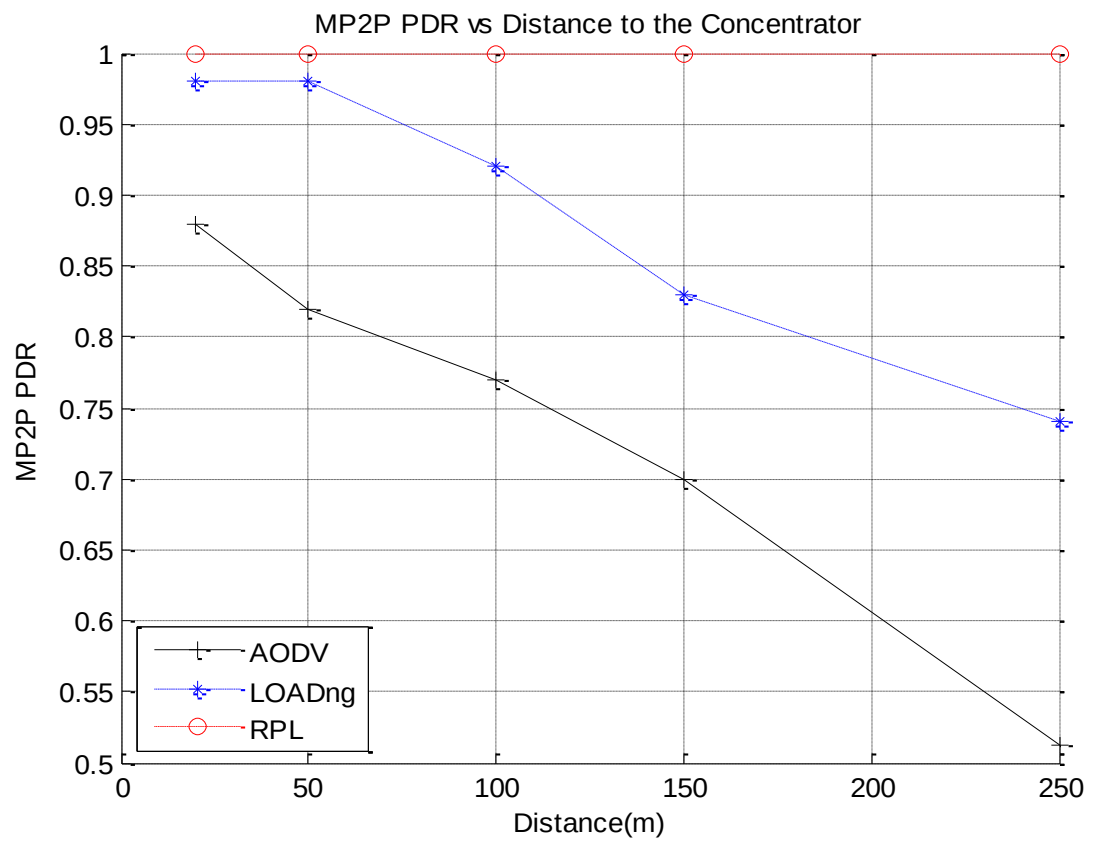

**Fig. 6. MP2P PDR function of Distance**

On the other hand, we observe a very low average latency in the case of standard RPL and it was equal to only 94 ms for both cases.

As LOADng's control packets are significantly smaller than those of AODV, LOADng control traffic overhead is considerably lower than AODV.

Fig.9 shows that a packet delivery ratio is 100% for RPL and the degradation for the delivered packets for AODV protocol in accordance with the increase of number of nodes; it reaches only 55% lower than LOADng which is equal to 75%. Whereas degradation for the delivered packets for AODV protocol in accordance with the distance to the concentrator reaches only 65% lower than LOADng which is equal to 85%.

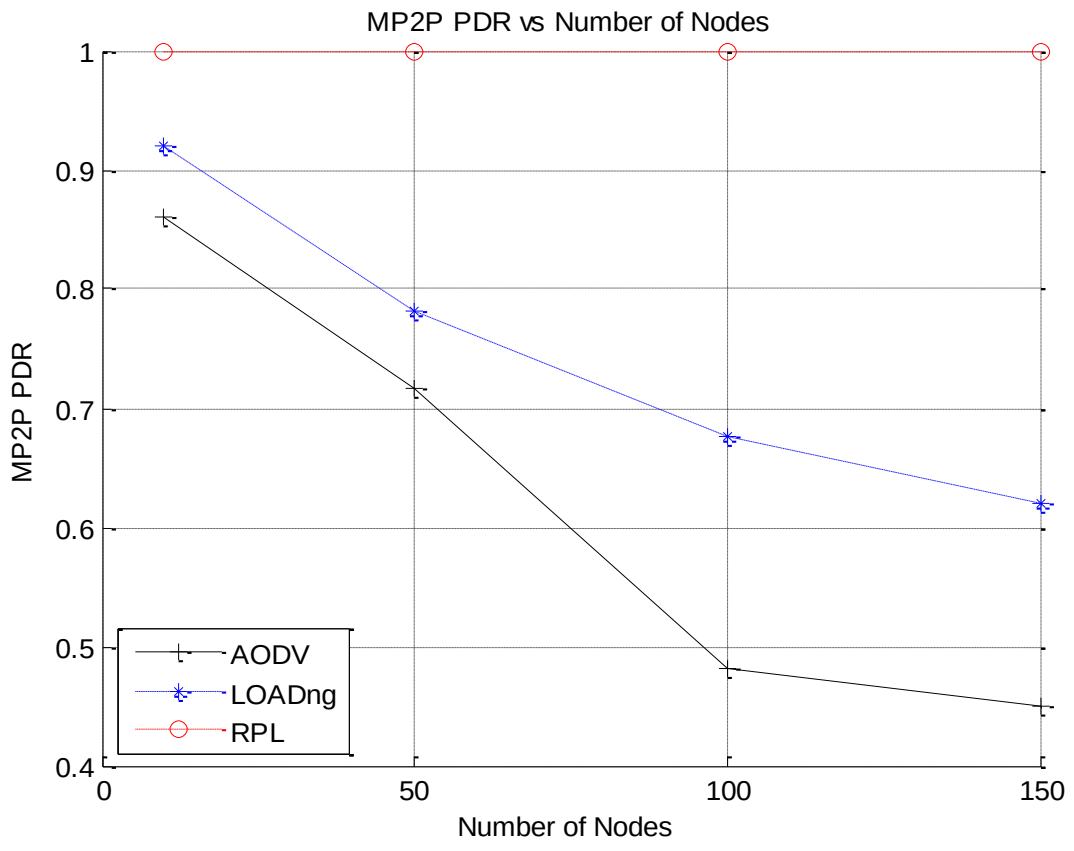

**Fig. 7. MP2P PDR function of Nodes Number**

For upward traffic, Fig.10 shows that LOADng is better than AODV in terms of average latency which is equal to 569 ms compared to AODV 2884 ms when the network is subject to variable distance to the concentrator and when the network is with increasing number of nodes is variable latency is up to 1062 ms and 3586 ms for LOADng and AODV respectively. On the other hand, we observe a very low average latency in the case of standard RPL is equal to only 96 ms for both cases.

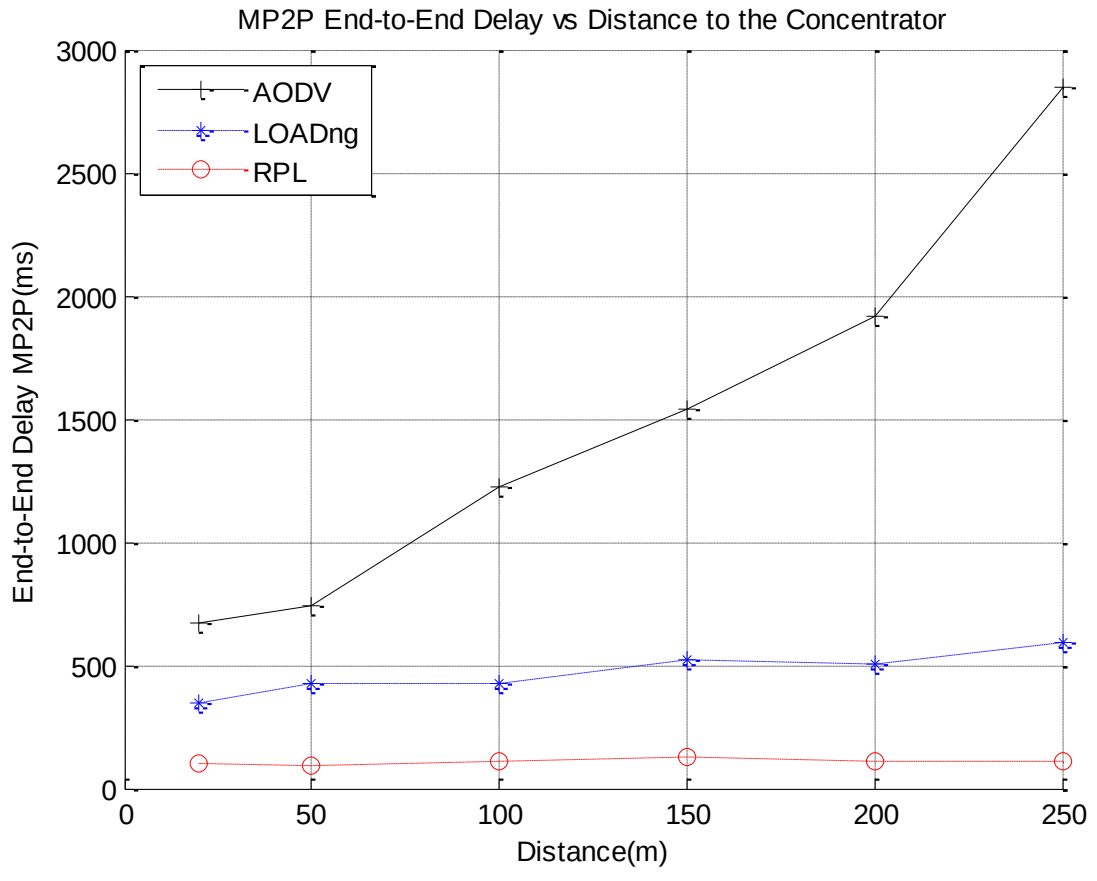

**Fig. 8. MP2P End-to-End Delay function of Distance**

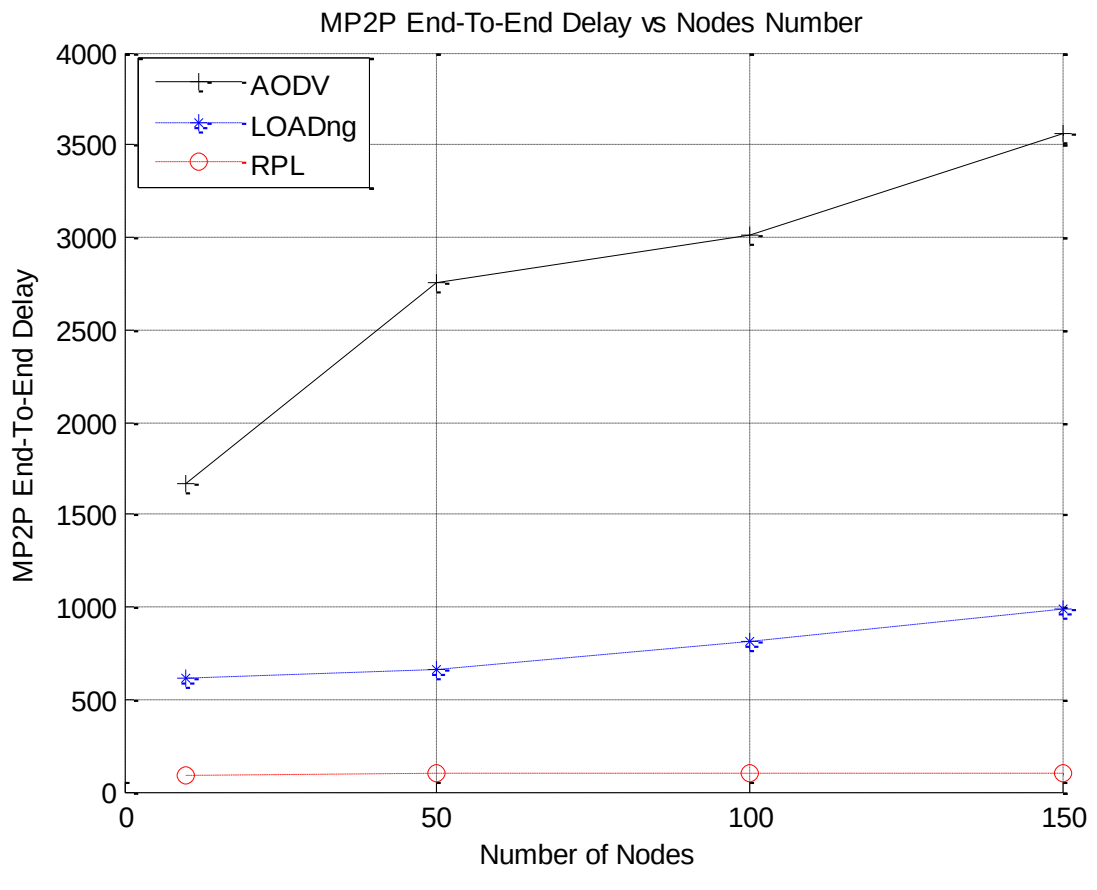

**Fig. 9. MP2P End-to-End Delay function of Node Number**

Fig.12 shows the number of overhead packets conducted by each router and Fig.13 depicts the Overhead bytes per second in the network incurring from respectively AODV, LOADng and RPL. LOADng protocol shows better performance than AODV in both cases.

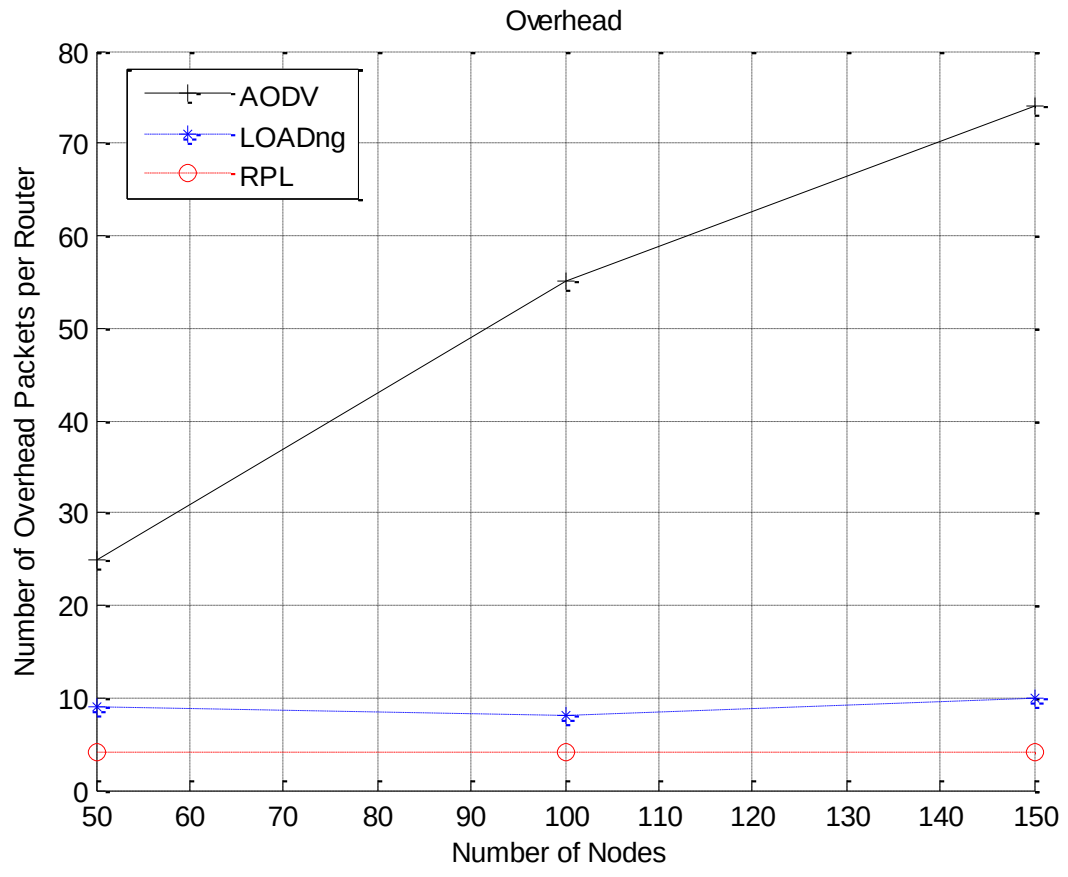

**Fig. 10. Number of overhead packets**

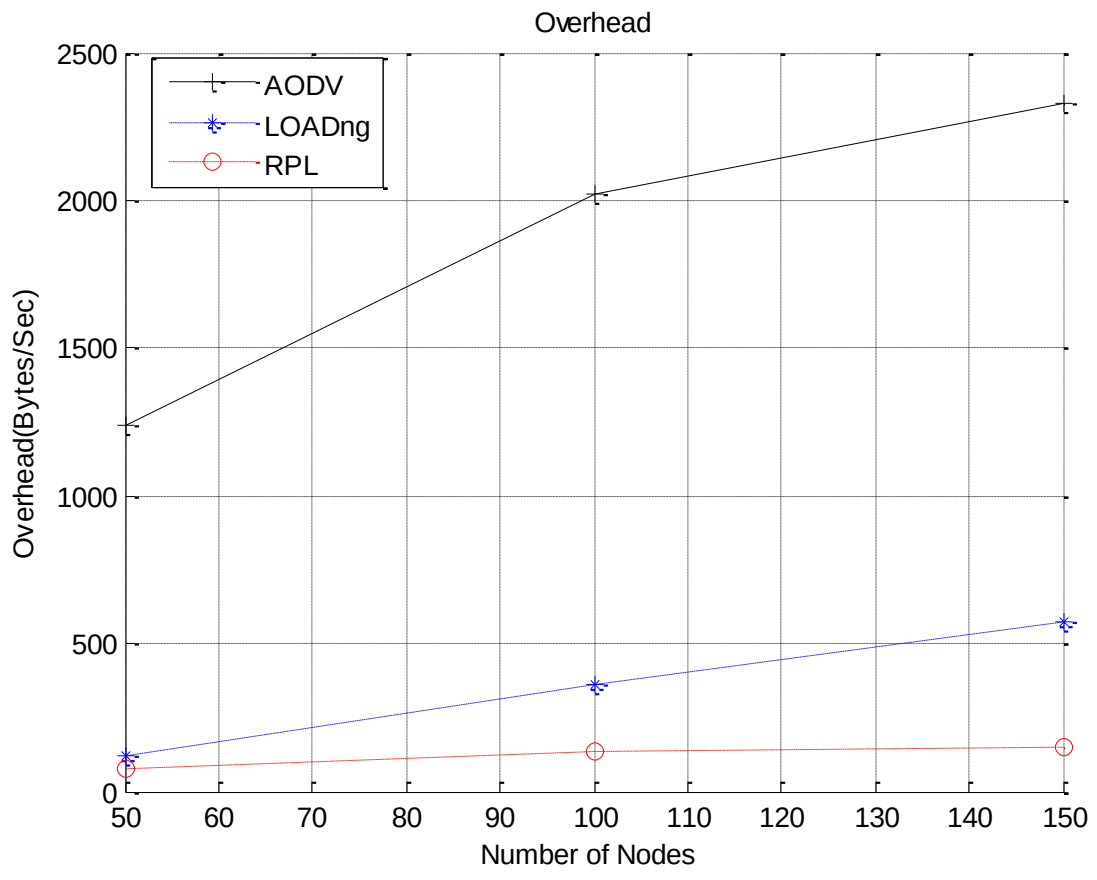

**Fig. 11. Overhead bytes per second**

Our study reveals that the LOADng implementation harvests better performance than AODV in all used metrics: higher data delivery ratios, lower delays and lower overhead.

In all cases RPL protocol performs better than our LOADng implementation but in other hand LOADng protocol represents a part of the ITU-T G.9903 recommendation for smart metering applications. Its strength came from its simple processing compared to RPL protocol, even though protocol extension should be developed to optimize its performance for AMI applications.

## IV. CONCLUSION

The results in simulations reveal that the LOADng protocol is better than AODV routing protocol in all cases. In AMI scenario, LOADng showed better memory management and power consumption. Also, it still has less implementation complexity compared to RPL which is a crucial point.

To sum up, many aspects remain interesting perspectives and future challenges: Evaluating LOADng performance in real experiment, testing its performance in other application domains in order to find how we can optimize flooding and route storage to maximize its performance.

## AUTHORS

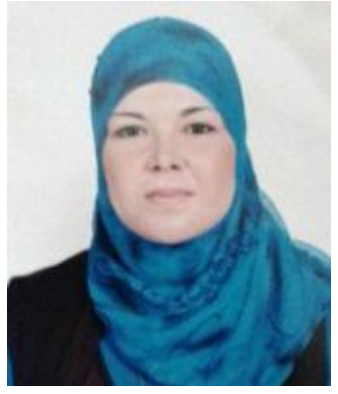

**S. Elyengui**: is a PhD student in the department of communication systems at Tunisian National School of Engineering, University of Tunis El Manar Tunisia. She is a researcher in the area of smart grid communication and networking, SG networks security, AMI applications and M2M communications. She received her Computer Networks Engineer Diploma and a Master degree in new Technologies of Communication and Networking in 2007 and 2011 respectively.

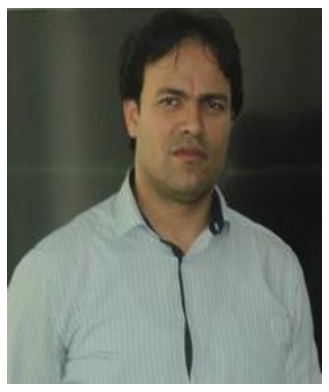

**R. Bouhouchi** received the Master degrees in communication systems from ENIT, the MBA degree s from Mediterranean School of Business, T.E graduated in 2000 from ESPTT, and holds an engineering degree in computer sciences since 2006, as he holds more than eight international certificates in advanced programming and management as ITIL (Information Technologies I